\documentclass[%
 reprint,
 amsmath,amssymb,
 aps,
]{revtex4-1}

\usepackage{graphicx}% Include figure files
\usepackage{dcolumn}% Align table columns on decimal point
\usepackage{bm}% bold math
\usepackage{color}
\DeclareMathOperator{\csch}{csch}

\newcommand{\be}{\begin{equation}}
\newcommand{\ee}{\end{equation}}
\newcommand{\bea}{\begin{eqnarray}}
\newcommand{\eea}{\end{eqnarray}}
\newcommand{\nn}{\nonumber}

\begin{document}

%\preprint{APS/123-QED}

%\title{Doping-dependent characteristic temperatures in two-dimensional Dirac materials}
\title{  Bloch-Gr\"{u}neisen temperature and universal scaling of normalized resistivity in doped graphene revisited}
\author{Khoe Van Nguyen$^{1,2,3}$}
\email{nvkhoe@gate.sinica.edu.tw}
\author{Yia-Chung Chang$^{1,4}$}
\email {yiachang@gate.sinica.edu.tw}
\address{$^{1}$ Research Center for Applied Sciences, Academia Sinica, Taipei 115, Taiwan}
\address{$^{2}$ Molecular Science and Technology, Taiwan International Graduate Program, Academia Sinica, Taipei 115, Taiwan}
\address{$^{3}$ Department of Physics, National Central University, Chungli, 320 Taiwan}
\address{$^{4}$ Department of Physics, National Cheng-Kung University, Tainan 701, Taiwan}

\date{\today}

\begin{abstract}

In this work, we resolved some controversial issues on the Bloch-Gr\"{u}neisen (BG) temperature in doped graphene via analytical and numerical calculations based on full inelastic electron-acoustic-phonon (EAP) scattering rate and various approximation schemes.  Analytic results for BG temperature obtained by semi-inelastic (SI) approximation (which gives scattering rates in excellent agreement with the full inelastic scattering rates) are compared with those obtained by quasi-elastic (QE) approximation and the commonly adopted value of $\Theta^{LA}_{F} = 2\hbar v_{LA} k_F/k_B$. It is found that the commonly adopted BG temperature in graphene ($\Theta^{LA}_{F}$) is about 5 times larger than the value obtained by the QE approximation and about  2.5 times larger than that by the SI approximation, when using the crossing-point temperature  where low-temperature and high-temperature limits of the resistivity meet (criterium 1). The corrected analytic relation based on SI approximation agrees extremely well  with the transition temperatures determined by  fitting the the low- and high-$T$ behavior of available experimental data of graphene's resistivity. We also introduce  a way to determine the BG temperature  including the full  inelastic EAP scattering rate and the deviation of electron energy from the chemical potential ($\mu$) numerically by finding  the maximum of  $\partial \rho(\mu,T)/\partial  T$ (criterium 2). It is found that the BG temperature determined by the full numerical calculation with  criterium 2 falls between the values obtained analytically via the SI appoximation with criterium 1 ($\Theta_{BG,1}$) and  criterium 2 ($\Theta_{BG,2} \approx 1.35\Theta_{BG,1}$) but neglecting the contribution of electron energies away from $\mu$.  Using the analytic expression of $\Theta_{BG,1}$ we can prove that the normalized resistivity  defined as $R_{1}=\rho(\mu,T)/\rho(\mu,\Theta_{BG,1})$ plotted as a function of $(T/\Theta_{BG,1})$ is independent of the carrier density. Applying our results to the experimental data extracted from [Phys. Rev. Lett. 105, 256805 (2010)] shows  a  universal scaling behavior, which is different from previous studies.

\end{abstract}


\maketitle

%\section{\label{sec:level1}Introduction}
%\section{\label{sec:level1}Introduction\protect}

Bloch-Gr\"{u}neisen  temperature ($\Theta_{BG}$) in doped graphene has been extensively discussed in the literature [1-31]. $\Theta_{BG}$ is  ususally defined as the crossing point in temperature between the low-temperature (LT) limit expression and high-temperature (HT) limit expression for resistivity as functions of temperature [1-10]. $\Theta_{BG}$ is an important characteristic temperature for designing graphene-based devices in applications such as optical detectors, bolometers \cite{Yan2012,Betz2013} and in cooling pathways and supercollisions \cite{Bistritzer2009,Betz2013,McKitterick2016,Song2012,Ma2014,Tikhonov2018,Kong2018}.
If one considers separate contributions in $\Theta_{BG}$ due to  longitudinal-acoustic (LA) and transverse-acoustic (TA) phonon scatterings, simple analytic expressions for  $\Theta_F^{a}$ ($a=LA,TA$) can be obtained based on the LT- and HT-limit expressions for resistivity.
The current general consensuses is that $k_B\Theta_F^a = 2\hbar v_a k_F$ ($a = LA,\,TA$) through analyses based on  quasi-elastic scattering approximation [1-27]. In the LT limit, it can be shown that $\rho^{LT}_a(\mu,T)\propto   T^4$ and in the HT regime $\rho^{HT}_a(\mu,T) \propto T$\cite{Hwang2008,Efetov2010}.

Critical inconsistencies exist in Figs.~2b and 3b of \cite{Efetov2010} and in Fig.~2 of \cite{DSouza2017}. The crossing points found in both studies are  $\sim$ 5 - 6 times smaller than the theoretical values predicted by $\Theta_F^{LA}$ above.  To remedy this inconsistency an artificial scaling parameter of $\zeta = 0.2$ was introduced in \cite{Efetov2010} to make the BG resistivity $\rho(\mu,\zeta\Theta_F^{LA})$ fall within the experimentally accessible range. Using the same definition of $\Theta_F^a$, it was shown that the BG transition occurs for $T < 0.2 \Theta_F^{LA}$ \cite{Yan2012}, $T \lesssim 0.15 \Theta_F^a$ \cite{Sohier2014}, $T \lesssim 0.25 \Theta_F^{a}$ \cite{McKitterick2016,Ansari2017}, and $1/\tau^{LT}_a(\mu,T) \propto T^4$ for $T \lesssim 0.2 \Theta_F^a$ as implied in Ref.~\cite{Mariani2010}. Our analysis based on quasi-elastic approximation also gives the same conclusion, i.e. $k_B\Theta_{BG,0} = 2\hbar v_{LA} k_F/5$. More detailed analyses are given in supplemental material (SM)\cite{Nguyen2020SM}.

Contrary to the above, the analyses in Refs.\cite{Song2012,Ma2014,Tikhonov2018,Kong2018} suggest that $k_B\Theta_{BG} = \hbar v_{LA} k_F$. Note that in Ref.~\cite{Betz2013} $\Theta_{BG}=\Theta_F^{LA} = 2\hbar v_{LA} k_F/k_B$ was  used in the arguments and calculations  but their results are compared not only with those from Refs.~\cite{Bistritzer2009,Kubakaddi2009,Viljas2010} using the same $\Theta_F^{LA}$ but also with those from Ref.~\cite{Song2012} using $\Theta_{BG} = \hbar v_{LA} k_F/k_B$. To resolve these controversies a more careful analysis of the BG temperature and a revisit of the universal scaling of the normalized resistivity $R(T/\Theta_{BG})$ in doped graphene is needed.

%\section{Theory}

The resistivity in doped graphene  can be calculated according to\cite{Ashcroft1976,Nguyen2020}
\be
\rho^{-1}(\mu,T) = \sigma(\mu,T) = {e^2} \int \frac{k dk}{\pi} v_F^2 \tau(\epsilon_{k}) [ - \frac{d f(\epsilon_{k})}{d\epsilon_{k}} ],\label{rho}
\ee
where  $\sigma$ is  the conductivity. In common practice, $-d f(\epsilon_k)/d\epsilon_k$  is approximated by $\delta(\epsilon_k-\mu)$ since $\tau(\epsilon_{k}) |\epsilon_{k}| $ is slow varying over the range of $k_BT$. With this approximation we have\cite{Nguyen2020}
\bea
&&\rho_{\mu}(\mu,T) =\frac{\vert\mu\vert}{4 e^2\hbar v_F^3 \rho_m} \nn \\
&&\int  d\theta ( 1-\cos \theta)  \sum_{a,p}  D^p_{a}(\theta) \csch(\hbar \Omega^p_{a}/k_BT),\label{rhomu}
\eea
where $D^p_{a}(\theta)\approx B^2 T^p_{a}(\theta)$ (when $E_1$ is neglected) and $T^p_{a}(\theta)$ describes the angular dependence of the net electron-acoustic-phonon (EAP) scattering strength with $a=LA,TA$ and $p=\pm$ for phonon absorption or emission ($T^p_{a}(\theta)$ is depicted in Fig.~1 of \cite{Nguyen2020}). $\hbar \Omega^p_{a}$ is the corresponding phonon energy.  Throughout the paper, we only consider the $n$-doped case. Due to electron-hole symmetry in the Dirac Hamiltonian, the behavior of $p$-doped case will be identical. We use $v_F = 1.0 \times 10^6$ (m/s), $v_{LA} = 2.0 \times 10^4$ (m/s), $v_{TA} = 1.3 \times 10^4$ (m/s), $\rho_m = 7.6 \times 10^{-7}$ (Kg/m$^2$) \cite{Nguyen2020,Kumaravadivel2019}, $g_0$ = 20 (eV) and $\beta = 3$ \cite{Castro2010,Nguyen2020}.

%$\Leftrightarrow 1/\tau(\mu,LT) = 1/\tau(\mu,HT)$ gives \cite{Nguyen2020SM}
Using the quasielastic (QE) approximation for  scattering rates \cite{Hwang2008,Efetov2010} and setting $\rho_{QE}^{LT}(\mu,T) = \rho_{QE}^{HT}(\mu,T)$  gives \cite{Nguyen2020SM}
\begin{eqnarray}
k_B\Theta_{BG,0} = \sqrt[3]{15} \hbar v_{LA}k_F /2\pi \approx k_B\Theta_F^{LA}/5,\label{TBG0}
\end{eqnarray}
where $\Theta_F^{LA} = 2\hbar v_{LA} k_F/k_B $ is a characteristic temperature, which is often used as the BG temperature in the literature [1-27].  The more appropriate BG temperature within quasielastic approximation should be $\Theta_{BG,0}$, although it is still quite different from the results derived from the full calculation.
%$k_B\Theta_0 = \sqrt[3]{15} \hbar v_{LA} |\mu|/2\pi \approx 0.4 \hbar v_{LA} |\mu| = {2\hbar v_{LA}k_F}/{5} = k_B\Theta_F^{LA}/5$ \cite{Nguyen2020SM}.
The transferred acoustic phonon energy is determined by $\hbar\omega_{a}^p =\hbar v_a q_{a}^p \approx 2 \hbar v_a k\sin(\theta/2)$ since $v_a/v_F \ll 1$  \cite{Nguyen2020}. Thus, we obtain $\left<\hbar\omega_{a}^p\right>_p = 2\hbar v_ak_F\sin(\theta/2)$ for $\epsilon_{k}=\mu$. Therefore, $ \max \left( \left<\hbar\omega_{a}^p\right>_p \right) = 2\hbar v_a k_F =k_B\Theta^a_{F}$ has the physical meaning of the maximal transferred acoustic phonon energy.
% we reproduce the result adopted in literature \cite{Hwang2008,Bistritzer2009,Kubakaddi2009,Efetov2010,Fuhrer2010,Viljas2010,Castro2010,Mariani2010,Min2011,Sarma2011,Yan2012,Chen2012,Cooper2012,Munoz2012,Fong2012,
%Somphonsane2013,Fong2013,Betz2013,Park2014,Sohier2014,McKitterick2016,Meunier2016,Ansari2017,Rani2017,DSouza2017,Gunst2017,Ansari2018}
\begin{eqnarray}
k_B\Theta^a_{F} =  2\hbar v_a k_F.\label{TBGa0}
\end{eqnarray}
In Refs.~\cite{Fuhrer2010,Betz2013} it is suggested that $k_B \Theta_{BG}$ is close to the maximum transferred phonon  energy.
%However, it is irrational to assume that only phonons with $\theta = \pi$ have the overwhelmingly dominating contribution. In fact, because of the prefactor $(1-\cos\theta)$ in the scattering rates \cite{Nguyen2020}, phonons with $\pi/2 < \theta < 3\pi/2$ also have significant contribution to carrier transport. Therefore, this  oversimplifications made $\Theta^a_{F}$ be $\sim 2-2.5$ times greater than our best estimated value of the BG temperature.

Next, we consider a semi-inelastic (SI) approximation, which gives results very close to the full inelastic scattering calculation \cite{Nguyen2020}. In the SI approximation, the LT and HT limits of $\tau^{-1}(\mu,T)$ are given  by
\begin{subequations}
\label{rho1}
\be
\rho_{SI}^{LT}(\mu,T)= \frac{3v_F(k_B T)^4}{e^2\rho_m|\mu|^3\hbar v_{LA}^5} \left[ 2E_1^2+B^2 (1 + \beta_v^5) \right]  \label{rho1a}
\ee
and
\be
\rho_{SI}^{HT}(\mu,T) = \frac{\pi k_BT}{4e^2\rho_mv_{LA}^2\hbar v_F^2} \left[ E_1^2 + 2B^2 (1 + \beta_v^2) \right], \label{rho1b}
\ee
\end{subequations}
respectively, where $E_1=g_0/\epsilon(q)$ is the screened deformation potential for $LA$ phonon, $B={3\beta\gamma_0}/{4}$ denotes electron-phonon coupling strength due to  unscreened gauge fields for both $LA$ and $TA$ phonons, and  $\beta_v = v_{LA}/v_{TA}$ with $v_{LA}\,(v_{TA})$ being the sound velocity of $LA\,(TA)$ phonon. Since $E_1^2/B^2 \ll 1$, it is a good approximation to neglect the $E_1$ contribution.

By setting $\rho_{SI}^{LT}(\mu,T) = \rho_{SI}^{HT}(\mu,T)$ we obtain  \cite{Nguyen2020SM}		
\begin{subequations}
\label{TBG1}
\begin{eqnarray}
\Theta_{BG,1} &=& \sqrt[3]{\frac{2E_1^2\Theta_{d,1}^3 + B^2 (\Theta_{LA,1}^3 + \beta_v^5 \Theta_{TA,1}^3)}{2E_1^2 + B^2 (1 + \beta_v^5)}},\label{TBG1d} \\
&\approx& \sqrt[3]{\frac{\Theta_{LA,1}^3 + \beta_v^5 \Theta_{TA,1}^3}{1 + \beta_v^5}} =\sum_a b_{a} \hbar v_{a} k_F /k_B, \label{TBG1e}
\end{eqnarray}
where
\bea
k_B\Theta_{d,1} &=& \sqrt[3]{{\pi}/{24}} \hbar v_{LA}k_F \approx {\hbar v_{LA}k_F}/{2}\label{TBG1a}\eea
and
\bea
k_B\Theta_{a,1} &=& \sqrt[3]{{\pi}/{6}} \hbar v_{a}k_F \approx {4\hbar v_{a}k_F}/{5} \label{TBG1b}
\eea
\end{subequations}
($a=LA,TA$) denote separate contributions from deformation potential (labeled by $d$) and unscreened LA and TA phonon scatterings.
And $b_{a} = \sqrt[3]{{\pi}/{6}}(\delta_{a,LA} + \beta_v\delta_{a,TA}) \sqrt[3]{\frac{1+\beta_v^2}{1+\beta_v^5}}$.

Because the  resistivity $\rho_{\mu}(\mu,T)$ is proportional to $T^4$ ($T$) in the low-$T$ (high-$T$) limit, $\Theta_{BG}$ should be near the maximum of $\partial \rho_{\mu}(\mu,T)/\partial T$. Using the semi-inelastic scattering rate at $\epsilon_{k} = \mu$, we have\cite{Nguyen2020}
\begin{eqnarray} && {\rho_{\mu}(\mu,T)} = \frac{3v_F(k_B T)^4}{e^2\rho_m\hbar|\mu|^3}  \nonumber \\
&& \left\{ \frac{1}{v_{LA}^5} \left[ \frac{2E_1^2}{1+c_1 \alpha_{LA}^3} +\frac {B^2} {1+ c_0\alpha_{LA}^3} \right] + \frac{1}{v_{TA}^5}\frac {B^2} {1+ c_0\alpha_{TA}^3} \right\} \nn \\
&& \approx \frac{\hbar\Lambda(k_B T)^4}{e^2 |\mu|^3} \sum_a \frac{\delta_{a,LA} + \beta_v^5\delta_{a,TA}}{1+ c_0\alpha_a^3} \label{SI}
\end{eqnarray}
where $\Lambda = \frac{3v_FB^2}{\rho_m\hbar^2v_{LA}^5}$,  $c_0 = 16.5,\,c_1 = 65.7$, and $\alpha_a = T/\Theta^a_{F}$ \cite{Nguyen2020}.

\begin{figure}[t!]
\includegraphics[width=1.\linewidth]{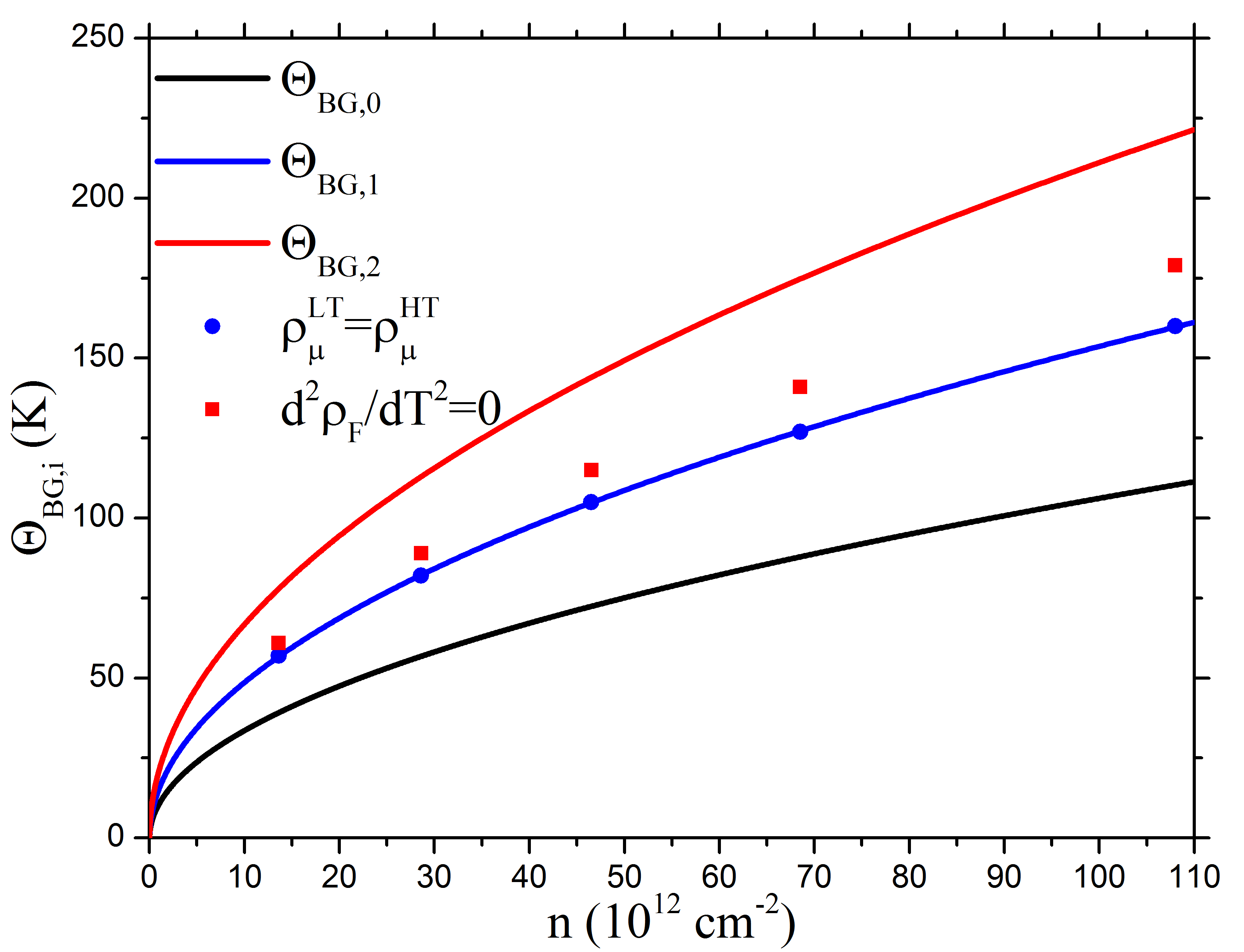}
\caption{\label{ms1} BG temperatures $\Theta_{BG,i}$ determined by three different ways with solid black, blue, and red curves for $i=$ 0, 1, and 2, respectively. Blue circles display  $\Theta_{BG,1}$ inferred from fitting LT and HT limits of experimental data of \cite{Efetov2010} at five densities ranging from $13.6 - 108 \times 10^{12} cm^{-2}$ by using scattering rates for $\epsilon_{k} = \mu$ with semi-inelastic approximation. The red squares are obtained by taking derivatives of the resistivity including contributions from all $\epsilon_{k}$'s with full inelastic scattering rates.}
\end{figure}

 Using Eq.~(\ref{SI}) we can solve the equation
 \be
 \partial ^2\rho_{\mu}(\mu,T)/\partial T^2 = 0 \label{rho2}
 \ee
analytically and get the BG temperature at the peak of $\partial \rho_{\mu}(T)/\partial T$ (See Sec. III in SM \cite{Nguyen2020SM} for derivations )

 \begin{subequations}
\label{TBG2}
\begin{eqnarray}
 \Theta_{BG,2} \approx \sqrt[3]{\frac{\Theta_{LA,2}^3 + \beta_v^2 \Theta_{TA,2}^3}{1 + \beta_v^2}} = \tilde b_{a} \hbar v_{a}k_F/k_B, \label{TBG2d}
\end{eqnarray}
where
\bea
k_B \Theta_{a,2} &=& \sqrt[3]{{16}/{c_0}} \hbar v_{a} k_F  \,\label{TBG2b}
\end{eqnarray}
\end{subequations}
with $\tilde b_{a} = (\delta_{a,LA} + \beta_v\delta_{a,TA}) \sqrt[3]{\frac{3(1+\beta_v^{-1})}{c_0(1+\beta_v^2)}}$. Note that the $E_1$ term has been neglected in Eq.~(\ref{TBG2d}).  We found
\be
\Theta_{BG,1} /\Theta_{BG,2} = b_{a} /\tilde  b_{a} \approx 0.728.
\ee
The theoretical results described by Eqs.~(\ref{TBG0}), (\ref{TBGa0}), (\ref{TBG1}), and (\ref{TBG2})  are plotted in Fig.~\ref{ms1} for comparison.
%Moreover, the experimentally determined results (below) are also included to validate our theoretical values.

Above we have shown that for the special case of $\epsilon_{k} = \mu$,  $\Theta_{BG,1}$  and $\Theta_{BG,2}$ obtained by two different approaches can differ by about 30\%. For full considerations including contribution from all possible $\epsilon_{k}$'s, it is impossible to determine the temperature dependence of resistivity analytically. However, it is possible to calculate $\rho(\mu,T) $ according to Eq.~(1) numerically with the full inelastic scattering rate without fixing $\epsilon_{k}$ at $\mu$. We then take the derivatives of the full $\rho(\mu,T) $ to determine $\Theta_{BG}$. The results are displayed by the red squares in Fig.~\ref{ms1} at five densities ranging from $13.6 - 108 \times 10^{12} cm^{-2}$ corresponding to samples studied in \cite{Efetov2010} and we see that $\Theta_{BG}$ so determined falls between  $\Theta_{BG,1}$ and $ \Theta_{BG,2}$.

\begin{figure}[h]
\includegraphics[width=1.\linewidth]{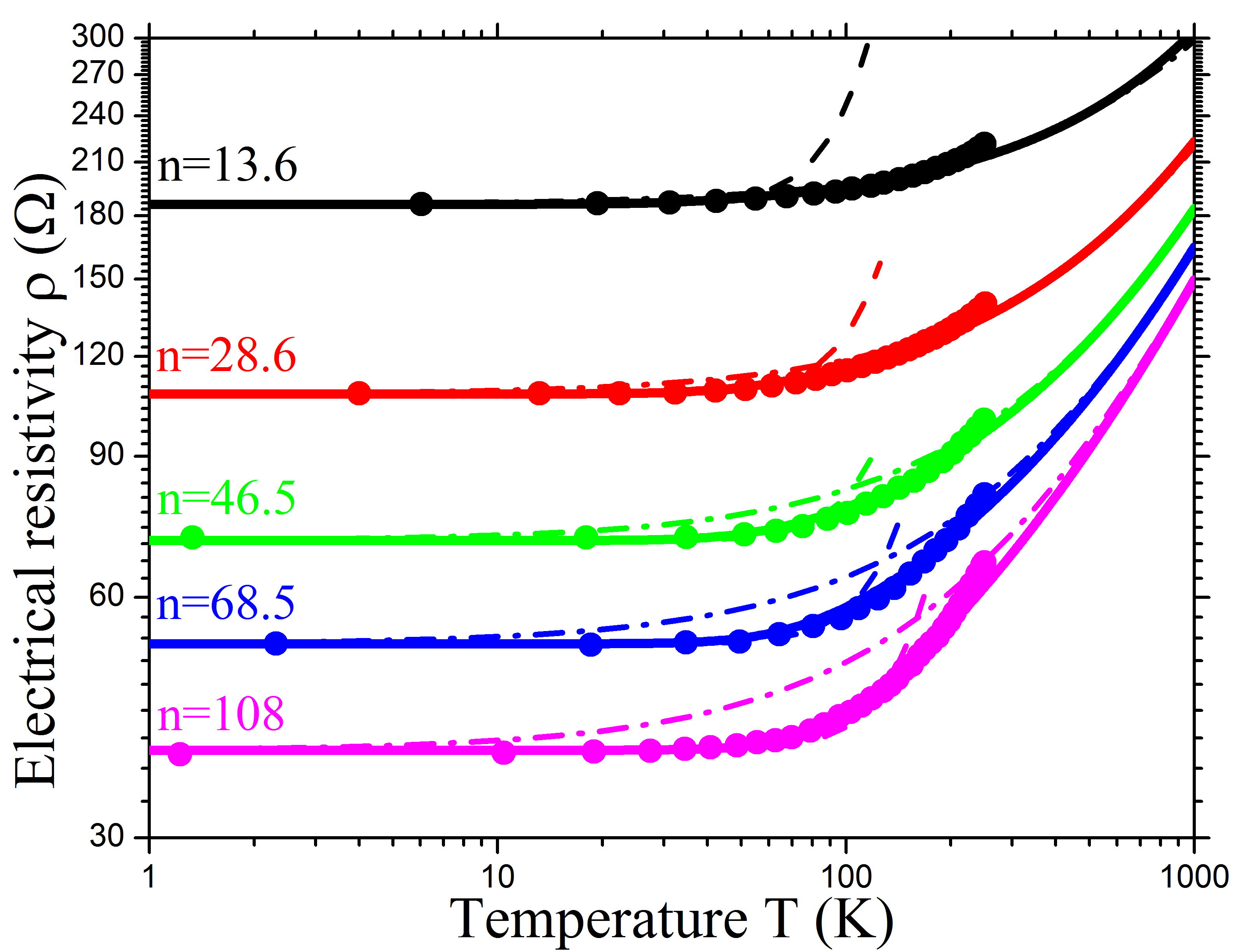}
\caption{\label{ms2}
Log-log plot of the electrical resistivity of graphene/SiO$_2$ with $n = 13.6 \times 10^{12}\,cm^{-2}$ (in black), $28.6 \times 10^{12}\,cm^{-2}$ (in red), $46.5 \times 10^{12}\,cm^{-2}$ (in green), $68.5 \times 10^{12}\,cm^{-2}$ (in blue), and $108 \times 10^{12}\,cm^{-2}$ (in magenta) as a function of $T$. The solid curves are $\rho_{F}(n,T)$ obtained by Eq.~(\ref{rho}) with suitable parameters. The dashed-dotted and dashed curves respectively demonstrate the fitted high-$T$ and low-$T$ resistivities by using $\rho_{SI}^{HT}(\mu,T)$ and $\rho_{SI}^{LT}(\mu,T)$ given by  Eq.~(\ref{rho1}). The experimental data shown by colored circles are extracted from \cite{Efetov2010}.}
\end{figure}

To compare with experimental results, we  extract the experimental data of resistivities in n-doped graphene from \cite{Efetov2010} at five carrier densities and plot them  as colored dots in Fig.~\ref{ms2} for $n = 13.6 \times 10^{12}\,cm^{-2}$ (black), $28.6 \times 10^{12}\,cm^{-2}$ (red), $46.5 \times 10^{12}\,cm^{-2}$ (green), $68.5 \times 10^{12}\,cm^{-2}$ (blue), and $108 \times 10^{12}\,cm^{-2}$ ( magenta). We can fit these data well by using Eq.~(1) with the full inelastic scattering rates as solid curves in Fig.~\ref{ms2} which essentially go through the data points with slight deviation at the high-temperature end. To fit the data, a residual scattering rate beyond the acoustic phonon scattering mechanisms is added for a given $n$; that is, $1/\tau_0$ = 9.4, 7.9, 6.6, 5.95, and 5.5 $THz$ for the five respective carrier densities. The values of other constants adopted ($E_1, B, \beta_v$) are the same as those used in \cite{Nguyen2020}. We can also fit these data at high- and low-$T$ limits by using Eq.~(\ref{rho1}) with semi-inelastic scattering rates as shown in dash-dotted and dashed curves, respectively. The crossing points between those curves determine $\Theta_{BG,1}$, which are shown as  blue circles in Fig.~\ref{ms1} and they fall  perfectly on the theoretical curve (blue solid). For comparison,  $\Theta_{BG,0}$ determined by using the low- and high-$T$ quasielastic scattering rates \cite{Hwang2008,Efetov2010} as given in  Eq.~(\ref{TBG0}) as a function of density is shown as the black solid curve.

\begin{figure}[h]
\includegraphics[width=1.\linewidth]{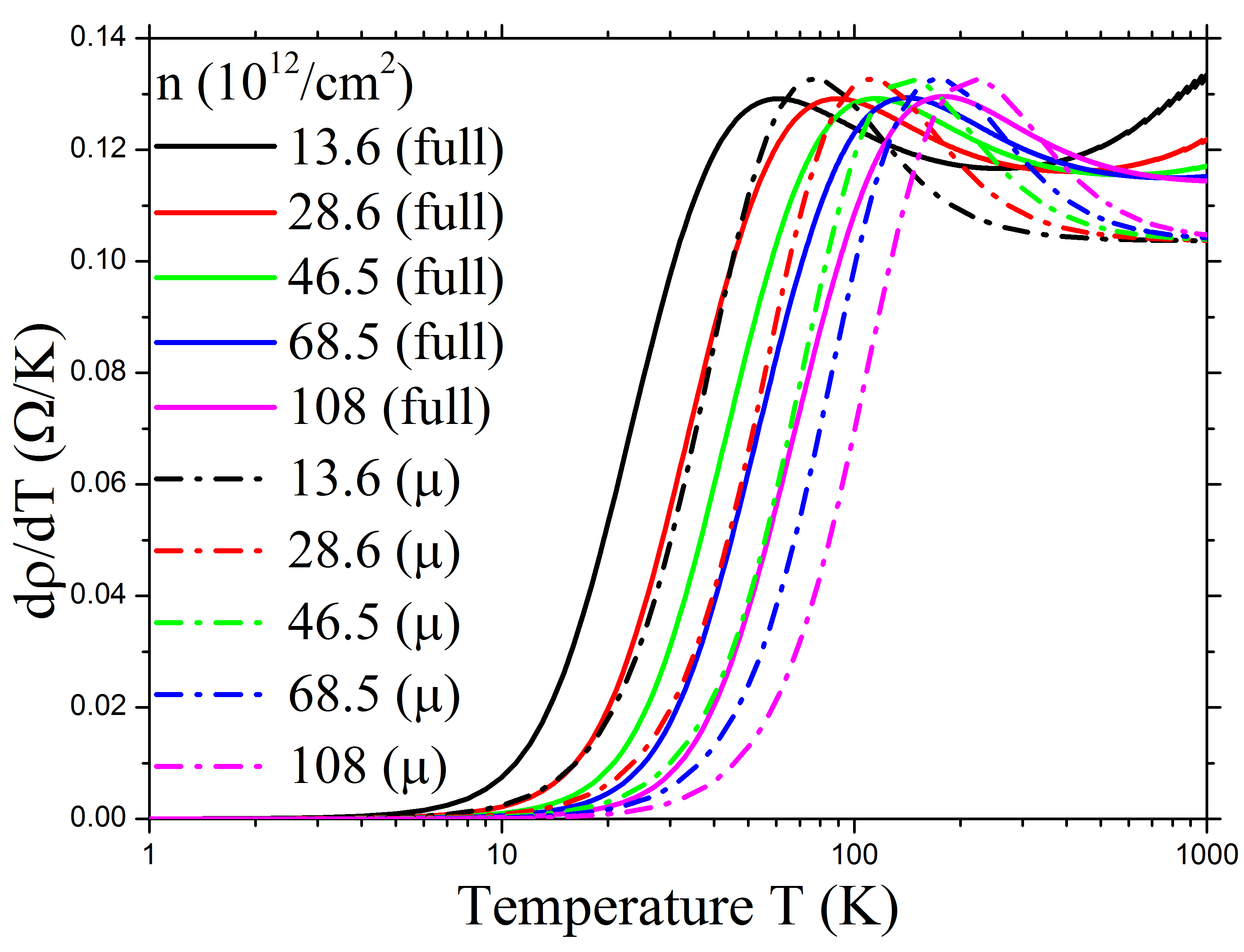}
\caption{\label{ms3}
Derivative of net resistivity,  $\partial \rho_{F}(n,T)/\partial T$ (which includes a constant correction term in scattering rate due to mechanisms beyond EAP scattering) as a function of $T$. Solid curves are from numerical calculation based on full inelastic scattering rate. For comparison, we also show $\partial \rho_{\mu}(n,T)/\partial T$ due to semi-inelastic scattering rate $\tau_{SI}^{-1}(\mu,T)$  (dashed-dotted curves)\cite{Nguyen2020}.}
\end{figure}

The total doping-dependent Bloch-Gr\"{u}neisen temperatures $\Theta_{BG}$ (shown by the red squares in Fig.~\ref{ms1} at five densities) are determined from the peak values of $\partial \rho_F(n,T)/\partial T$  plotted in Fig.~\ref{ms3}. Note that, $\rho_{F}(n,T)$ are obtained by using the full calculation described in Eq.~(\ref{rho}) with the inelastic scattering rate plus a correction term  ${1}/{\tau_0}$ which takes into account scattering mechanisms beyond the acoustic-phonon scattering. For comparison, we also show $\partial \rho_{\mu}(n,T)/\partial T$ as dash-dotted curves in Fig.~\ref{ms3}.
It should be noted that adding a constant correction term $1/\tau_0$ to the scattering rate will have no effect on  $\partial \rho_{\mu}(n,T)/\partial T$.
%In the inset, we remove $1/\tau_0$ and calculate the first derivatives of the EP-induced resistivity by using Eq.~(\ref{rho}) (the solid curves) and $\rho_{\mu}(\mu,T)$ with the scattering rate $1/\tau(\mu,T)$ above \cite{Nguyen2020} (the dashed-dotted curves). The peaks of the solid curves determine $\Theta$'s (which stay almost unchanged) and the maximum values of the dashed-dotted curves determine $\tilde \Theta_{BG}$'s described by the five red circles in Fig.~\ref{ms1}, which perfectly fall on the theoretical red solid curve by Eq.~(\ref{TBG2d}).
For a given $n$, $\Theta_{BG,2}$ shifts to the right of $\Theta_{BG}$ since the ratio of $\rho_{\mu}(n,T)$ to full $\rho(n,T) $ falls between $0.7$ and $1$\cite{Nguyen2020}. This comes from the fact that $-df(\epsilon)/d\epsilon$ can be approximated by $\delta(\epsilon-\mu)$ to obtain $\rho_{\mu}(n,T)$ only when $|\epsilon|\tau(\epsilon_{k})$ is slow varying over $ k_BT$, which is not quite satisfied in graphene \cite{Nguyen2020}.

Finally, we consider the universal scaling of normalized resistivity by using the well justified SI approximation and the full inelastic scattering rate for $\epsilon_k=\mu$. The normalized resistivity is defined as\cite{Ziman}	%,Efetov2010
\be
R_{i}(T/\Theta_{BG,i}) = {\rho(\mu,T)}/{\rho(\mu,\Theta_{BG,i})}.
\ee
In the SI approximation, $\rho(\mu,T)$ is given by Eq.~(\ref{SI}) and we have
\be\label{Rseti}
R_i\left( \frac{T}{\Theta_{BG,i}} \right) = \left(\frac{T}{\Theta_{BG,i}}\right)^4 \frac{ \sum_a \frac{\delta_{a,LA} + \beta_v^5\delta_{a,TA}}{1+ c_0\alpha_a^3} }
{ \sum_a \frac{\delta_{a,LA} + \beta_v^5\delta_{a,TA}}{1+ c_0\alpha_{a,i}^3} },
\ee
where $\alpha_{a,i} = \Theta_{BG,i}/\Theta^a_{F}$ with $i=1,2$.

From Eq.~(\ref{rho1}), we see that
$\rho_{SI}^{LT}(\mu,\Theta_{BG,i}) \propto \frac{(k_B\Theta_{BG,i})^4}{|\mu|^3}  \propto |\mu|$ and similarly
$\rho_{SI}^{HT}(\mu,\Theta_{BG,i}) \propto {(k_B\Theta_{BG,i})}  \propto |\mu|$. Therefore, the normalized resistivity of doped graphene is independent of $|\mu|$ and that is why one can get a universal curve for $R_{i}(T/\Theta_{BG,i})$ regardless of doping level of the sample.  This feature comes out naturally from our approach by using the semi-inelastic scattering rate. The normalized resistivity
$R_i$ is proportional to $\left({T}/{\Theta_{BG,i}}\right)^4 $ in the LT  limit and ${T}/{\Theta_{BG,i}}$ in the HT  limit.

Now instead of the SI approximation, we use the resistivity given in Eq.(\ref{rhomu}) with full inelastic scattering rate%, we obtain

\be\label{RinBG}
R^{in} = \frac{\int d\theta\sin^2\frac{\theta}{2} \sum_{a,p} \frac{T_{a}^p(\theta)}{v_a} \csch(\frac{Q_{a}^p(\theta){\Theta^a_F}}{2 T})}
{\int d\theta\sin^2\frac{\theta}{2} \sum_{a,p} \frac{T_{a}^p(\theta)}{v_a} \csch(\frac{Q_{a}^p(\theta)\Theta^a_{F}}{2 \Theta_{BG,i}})},
\ee
where $Q_{a}^p(\theta)=q_{a}^p/k$ is the ratio of the phonon momentum to electron momentum\cite{Nguyen2020}.
Our numerical results indicate that $R^{in}$ is almost identical to $R_1$ and $R_2$.
Obviously, as $T \rightarrow \Theta_{BG,i}$, $R_{i}$ and $R^{in}$ should approach 1.

\begin{figure}[t!]
\includegraphics[width=1.\linewidth]{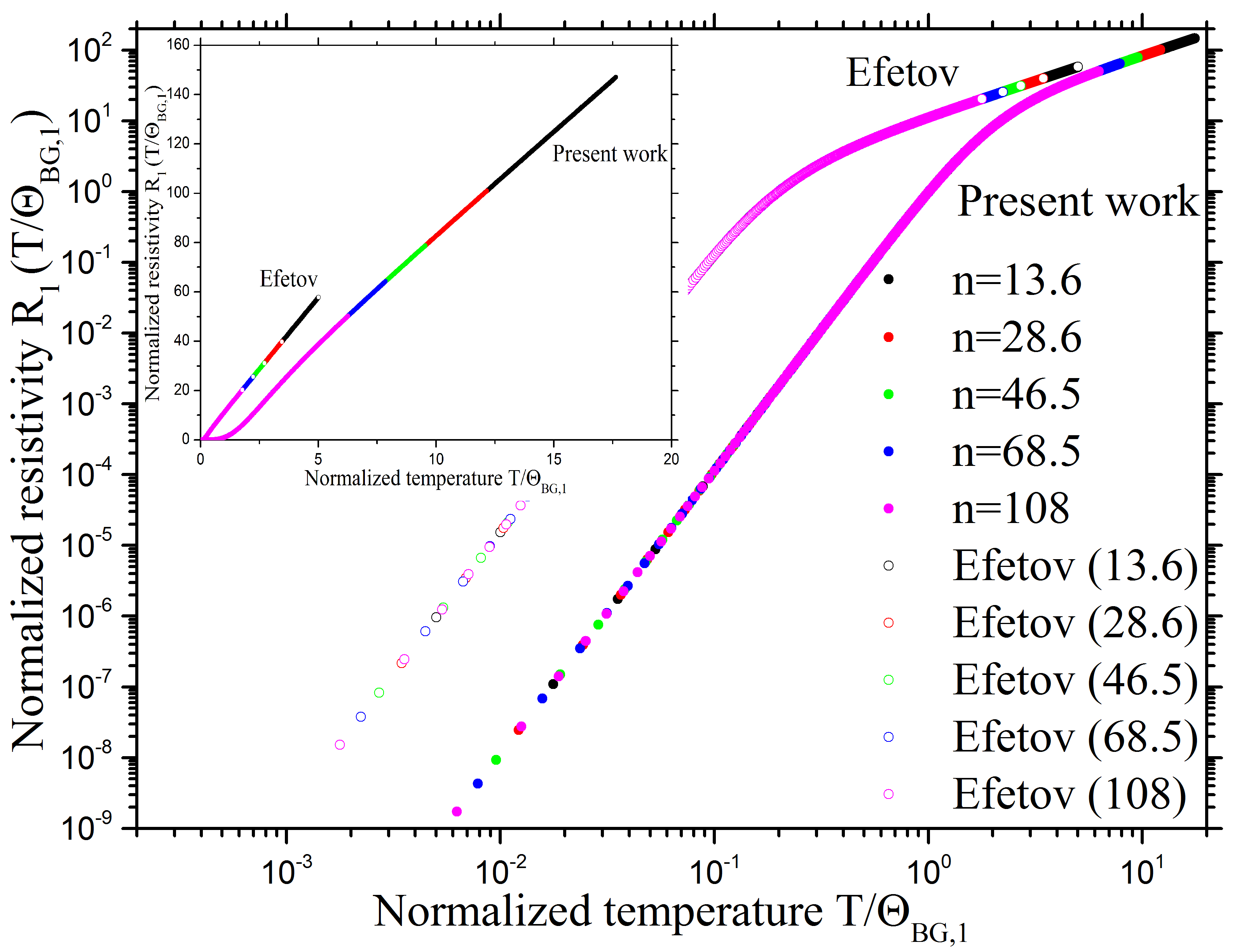}
\caption{\label{ms4}Log-log plot of normalized resistivity $R_{1}(T/\Theta_{BG,1})$ as a function of $T/\Theta_{BG,1}$ for various carrier densities ($n = 13.6-108 \times 10^{12}\,cm^{-2}$). The results of Ref.~\cite{Efetov2010} are also reproduced for comparison. The inset shows the corresponding linear plot of $R_{1}$. For ease of observing the universal behaviors, $T_{max}$ = 1000 K is used. }
\end{figure}

The results for $R_1$  together with the results taken from  Ref.~\cite{Efetov2010} are demonstrated in Fig.~\ref{ms4}. We found a significant difference between our results and those from \cite{Efetov2010}, especially for $T/\Theta_{BG,i} <1$. It is noted that in Ref.~\cite{Efetov2010} the normalized resistivity is defined as $R_0\left( \frac{T}{\Theta_{BG}} \right) ={\rho(\mu,T)}/{\rho(\mu,\xi \Theta_{BG})}$ with $\xi=0.2$ instead of 1. Since $\Theta_{BG}$ adopted in Ref.~\cite{Efetov2010} is  $2\hbar v_{LA} k_F/k_B=\Theta^{LA}_F$, which makes $0.2\Theta^{LA}_F\approx \Theta_{BG,0}$, the BG temperature determined by QE approximation given in Eq.~(\ref{TBG0}).
Although the universal scaling or behavior of $R_i$ as a function of the normalized temperature $T/\Theta_{BG,i}$ does not depend on $\mu$, for the same $T$ range of investigation, the heavier graphene gets doped (i.e. the larger $|\mu|$ induces the larger $\Theta_{BG,i}$), the narrower the range of $T/\Theta_{BG,i}$ becomes.
%For undoped  graphene, the electrical resistivity is linear in $T$ \cite{Nguyen2020}, $\rho_(0,T)$ has zero curvature (i.e. $d^2\rho_{(a)}(0,T)/dT^2=0$, which gives $\Theta_{BG,i}=0$ as implied in Eqs.~(\ref{TBG1}) and (\ref{TBG2}) for $\mu = 0$. And no universal scaling exists since $\rho (0,0)$ is undetermined.

%\section{Conclusion}

In conclusion, we have clarified the issues of BG temperatures in graphene via analytical and numerical calculations based on full inelastic EAP scattering rate and various approximation schemes.  We found that the commonly adopted BG temperatures in graphene ($k_B\Theta_F^{LA} = 2\hbar v_{LA} k_F$)
\cite{Hwang2008,Bistritzer2009,Kubakaddi2009,Efetov2010,Fuhrer2010,Viljas2010,Castro2010,Mariani2010,Min2011,Sarma2011,Yan2012,Chen2012,Cooper2012,Munoz2012,Fong2012,
Somphonsane2013,Fong2013,Betz2013,Park2014,Sohier2014,McKitterick2016,Meunier2016,Ansari2017,Rani2017,DSouza2017,Gunst2017,Ansari2018}
need to be corrected by a factor around 2.5, when using the same criterium [$\rho^{LT}_{\mu}(\mu,T) = \rho^{HT}_{\mu}(\mu,T)$]. The  BG temperature induced by the in-plane EAP scattering in semi-inelastic approximation is uncovered as $\Theta_{BG,1} \approx [(\Theta_{LA,1}^3 + \beta_v^5 \Theta_{TA,1}^3)/(1 + \beta_v^5)]^{1/3}$ with $\Theta_{a,1}=(\pi/6)^{1/3}\hbar v_a k_F/k_B$.  The corrected analytic relation agrees extremely well  with the transition temperatures determined by  fitting the the low- and high-$T$ behavior of available experimental data of graphene's resistivity \cite{Efetov2010}. We also show that Refs.~\cite{Mariani2010,Yan2012,Sohier2014,McKitterick2016,Ansari2017} well agree with the quasi-elastic (QE) prediction. When the  inelastic EAP scattering rate and the deviation of electron energy from the chemical potential ($\mu$) are fully taken into account, the resistivity $\rho(\mu,T)$ can only be described numerically. For this case we determine the  BG temperature by the point where $\partial \rho(\mu,T)/\partial  T$ is a maximum and thus $\partial^2\rho(\mu,T)/\partial T^2 = 0$ (criterium 2). If we also apply  criterium 2 to find the BG temperature in the SI approximation, we get  $\Theta_{BG,2} \approx [(\Theta_{LA,2}^3 + \beta_v^2 \Theta_{TA,2}^3)/(1 + \beta_v^2)]^{1/3}$ with $k_B \Theta_{a,2} = (16/c_0)^{1/3} \hbar v_a k_F$, which happen to be very close to the value $\hbar v_a k_F$ deduced in Refs.~\cite{Song2012,Ma2014,Tikhonov2018,Kong2018}.   We found that the BG temperature determined by the full numerical calculation with criterium 2 falls between the values obtained via the SI appoximation with criterium 1 ($\Theta_{BG,1}$) and  criterium 2 ($\Theta_{BG,2} \approx 1.35\Theta_{BG,1}$).  These values are about a factor 2 higher than the BG temperature ($\Theta_{BG,0}$) obtained with the  oversimplified QE approximation and a factor 2-2.5 lower than the commonly adopted value of $\Theta^{LA}_{F}$.

Finally, the resistivity normalized to its value at $T=\Theta_{BG,i}$ [$R_{i}=\rho(\mu,T)/\rho(\mu,\Theta_{BG,i})$] plotted as a function of the normalized temperature $T/\Theta_{BG,i}$ displays a  universal scaling behavior, which is independent of the carrier density\cite{Efetov2010}. Applying our results to the experimental data extracted from  Ref.~\cite{Efetov2010} does  show such a  universal scaling behavior, which obeys the relation $R_{i}(1) =1$.

\begin{acknowledgments}

Work supported in part by  Ministry of Science and Technology (MOST), Taiwan  under contract nos. 107-2112-M-001-032 and 108-2112-M-001-041.

\end{acknowledgments}

%\mbox{}\\ ${}^{\dagger}$ nvkhoe@gate.sinica.edu.tw\\
%\email[Corresponding author E-mail:] {yiachang@gate.sinica.edu.tw}

%\nocite{*}

%\bibliography{nvk1ref}% Produces the bibliography via BibTeX.

\end{document}

%\preprint{APS/123-QED}

%\title{Doping-dependent characteristic temperatures in two-dimensional Dirac materials}
\title{Supplemental material of\\ "Bloch-Gr\"{u}neisen temperature and universal scaling of normalized resistivity in doped graphene revisited"}

\author{Khoe Van Nguyen$^{1,2,3}$}
\email{nvkhoe@gate.sinica.edu.tw}
\author{Yia-Chung Chang$^{1,4}$}
\email {yiachang@gate.sinica.edu.tw}
\address{$^{1}$ Research Center for Applied Sciences, Academia Sinica, Taipei 115, Taiwan}
\address{$^{2}$ Molecular Science and Technology, Taiwan International Graduate Program, Academia Sinica, Taipei 115, Taiwan}
\address{$^{3}$ Department of Physics, National Central University, Chungli, 320 Taiwan}
\address{$^{4}$ Department of Physics, National Cheng-Kung University, Tainan 701, Taiwan}

\date{\today}


\begin{abstract}

In this supplemental material, we give a more detailed review of  current issues related to the Bloch-Gr\"{u}neisen temperature and  provide detailed derivations of all expressions discussed in the main text.

\end{abstract}

\maketitle

\section{\label{sec:level1}An overview of current issues related to doping-dependent Bloch-Gr\"{u}neisen temperatures in graphene}

Finite-temperature ($T$) vibrations of constituent components  in a crystal lattice produce quasiparticles named phonons $-$ quantum states of lattice vibrations, which in turn scatter off conducting charged carriers in the lattice causing electrical resistivity $\rho$ \cite{Ziman1960}. Quantum mechanically, the momentum and energy conservation laws must be fulfilled in these quantum processes \cite{Ziman1960}. In general, typical three-dimensional (3D) metals have large Fermi surfaces, reasonable Debye temperatures $\Theta_D$'s (i.e. transition temperatures between the high- and low-$T$ regimes of electron-phonon scatterings), and $\rho(T \gg \Theta_D) \propto T/\Theta_D$ and $\rho(T \ll \Theta_D) \propto (T/\Theta_D)^5$ \cite{Bloch1930,Gruneisen1933,Ziman1960}. Unlike conventional 3D metals \cite{Fuhrer2010}, graphene is a two-dimensional (2D) semimetal with zero bandgap, whose low-energy quasiparticles are described by a Dirac-like Hamiltonian with $\epsilon_{k}=\pm \hbar v_Fk$ ($+$ for n and $-$ for p-type). Thus, the Fermi surface of  doped graphene is a  small circle, which shrinks into a point in the intrinsic (undoped) limit. Moreover, since graphene has very strong in-plane $sp^2$ bondings resulting in an unusually high Debye temperature ($\Theta_D >$ 2000 K) \cite{Tewary2009}, new characteristic temperatures called Bloch-Gr\"{u}neisen (BG) temperatures $\Theta_{BG} \ll \Theta_D$ emerge. These new characteristic temperatures were also investigated in the low-$T$ transport of 2D parabolic systems in doped GaAs-Al$_x$Ga$_{1-x}$As \cite{Stormer1990} and GaAs-AlGaAs \cite{Raichev2017} heterostructures.

In 2008, Hwang \textit{et al.} \cite{Hwang2008} proposed a deformation-potential model, in which longitudinal acoustic ($LA$) phonons are assumed to dominate graphene's resistivity, and defined the BG temperature in doped graphene  as $k_B\Theta_{LA} = 2\hbar v_{LA} k_F$. Two years later, Efetov \textit{et al.} \cite{Efetov2010} reported that they had determined $k_B\Theta_{LA} = 2\hbar v_{LA} k_F$ experimentally by using the quasi-elastic scattering rates with $T$ and $T^4$ dependences in the high- and low-$T$ limits, respectively to fit their experimental data of graphene's resistivity at ultrahigh carrier densities.

However, recent theoretical and experimental works \cite{Castro2010,Sohier2014,Park2014,Li2014,Kumaravadivel2019,Greenaway2019,Nguyen2020} have shown that, in sharp contrast to Refs.~\cite{Hwang2008,Efetov2010}, transverse acoustic ($TA$) phonons contribute $\sim$ 2 times greater than $LA$ phonons. The new model is comprised of the screened deformation potential for $LA$ phonons, which only give a tiny contribution, while  the unscreened acoustic gauge fields for both $LA$ and $TA$ phonons give dominating contributions.

Therefore, there have so far existed some controversies in the literature  regarding the  BG temperature in graphene.

\section{Determinining Bloch-Gr\"{u}neisen temperatures by setting $\rho^{LT}(\mu) = \rho^{HT}(\mu)$}

From the electrical conductivity of n-doped graphene with $\epsilon_{k} = \mu$ at finite $\mu$ and $T$, $\sigma_{\mu}(\mu,T) = \frac{e^2}{\pi\hbar^2} \vert\mu\vert \tau(\mu,T)$, the electrical resistivity is calculated by $\rho_{\mu}(\mu,T) ={\sigma_{\mu}(\mu,T)}^{-1} = \frac{\pi\hbar^2}{e^2\vert\mu\vert} {\tau^{-1}(\mu,T)}$.
{In the quasielastic (QE) approximation (i.e. $\frac{1 - f(\epsilon')}{1 - f(\epsilon)} = 1$)} as adopted in References \cite{Hwang2008,Efetov2010}, the high-temperature (HT) and low-teperature (LT) quasielastic scattering rates are given by
\begin{eqnarray}
\frac{1}{\tau_{QE}^{HT}} &=& \frac{1}{\hbar^3}\frac{|\mu|}{4v_F^2}\frac{J_a^2}{\rho_m v_{LA}^2} k_BT,\label{tauHT0} \\
\frac{1}{\left<\tau_{QE}^{LT}\right>} &\approx& \frac{1}{\pi}\frac{1}{|\mu|}\frac{1}{k_F}\frac{J_a^2}{2\rho_m v_{LA}} \frac{4!\zeta(4)}{(\hbar v_{LA})^4} (k_BT)^4.\label{tauLT0}
\end{eqnarray}

Setting $\frac{1}{\tau_{QE}^{HT}} = \frac{1}{\left<\tau_{QE}^{LT}\right>}$ (i.e. $\rho_{QE}^{HT} = \rho_{QE}^{LT}$) leads to
\begin{eqnarray}
\frac{4!\zeta(4)}{\pi} \frac{v_F}{v_{LA}^3\mu^2} (k_B\Theta_{BG,0})^3 = \frac{1}{2} \frac{|\mu|}{v_F^2},
\end{eqnarray}
which gives $(k_B\Theta_{BG,0})^3 = \frac{\pi}{4!2\zeta(4)} (\hbar v_{LA} k_F)^3 = \frac{15}{8\pi^3} (\hbar v_{LA} k_F)^3$, where we have used $\zeta(4) = \pi^4/90$. Thus, we obtain
\begin{eqnarray}
k_B\Theta_{BG,0} &=& \frac{\sqrt[3]{15}}{2\pi} \hbar v_{LA} k_F
\approx 0.4 \hbar v_{LA} k_F = \frac{k_B\Theta_F^{LA}}{5},\label{TBG0}
\end{eqnarray}
where $\Theta_F^{LA}=2\hbar v_{LA} k_F $ is  a characteristic temperature, which is often interpreted as the Bloch-Gr\"{u}neisen temperature in the literature \cite{Hwang2008,Efetov2010,Castro2010,Sohier2014}.

Using the semi-inelastic scattering rates at HT and LT limits as discussed in \cite{Nguyen2020},
\begin{eqnarray}
\frac{1}{\tau_{SI}^{HT}(\mu,T)} &=& \frac{|\mu| k_BT}{4\rho_mv_{LA}^2\hbar^3v_F^2} \left[ E_1^2 + 2B^2 (1 + \beta_v^2) \right],\label{tauHT1} \\
\frac{1}{\tau_{SI}^{LT}(\mu,T)} &=& \frac{3v_F(k_B T)^4}{\pi\rho_m\mu^2\hbar^3v_{LA}^5} \left[ 2E_1^2+B^2 (1 + \beta_v^5) \right],\label{tauLT1}
\end{eqnarray}
we get
\begin{eqnarray}
E_1^2 + 2B^2 (1 + \beta_v^2) &=& \frac{12 T^3}{\pi(\hbar v_{LA}k_F)^3} \left[ 2E_1^2+B^2 (1 + \beta_v^5) \right],
\end{eqnarray}
at $T=\Theta_{BG,1}$, from which we obtain
\begin{eqnarray}
k_B\Theta_{BG,_1} &=& \sqrt[3]{\frac{\pi}{12} \frac{E_1^2 + 2B^2 (1 + \beta_v^2)}{2E_1^2+B^2 (1 + \beta_v^5)}} \hbar v_{LA} k_F \\
&=& \sqrt[3]{\frac{2E_1^2 \left(k_B\Theta_{d,1}\right)^3 + B^2 \left[ \left(k_B\Theta_{LA,1}\right)^3 + \beta_v^5 \left(k_B\Theta_{TA,1}\right)^3 \right]}
{2E_1^2 + B^2 (1 + \beta_v^5)}},
\end{eqnarray}
where
\[ k_B\Theta_{d,1} = \sqrt[3]{\frac{\pi}{24}} \hbar v_{LA} k_F, \]
and
\[ k_B\Theta_{a,1} = \sqrt[3]{\frac{\pi}{6}} \hbar v_{a} k_F \, (a=LA,TA) \label{LTA1} \]
are separate contributions to the BG temperature due to the negligibly screened deformation potential $E_1$ and the unscreened acoustic gauge field for ${LA}$ and ${TA}$ phonons. $\beta_v = {v_{LA}}/{v_{TA}}\approx 1.54 $ \cite{Kumaravadivel2019,bv}.

Since $E_1(q) = \frac{g_0}{\epsilon(q)}$ with $\epsilon(q) = \epsilon_r + \frac{g_sg_ve^2}{\hbar v_F} \frac{k_F}{q}$ and $B = \frac{3\beta\gamma_0}{4}$ with $\gamma_0 = \frac{2\hbar v_F}{\sqrt{3}a_0}$ and $a_0 = \sqrt{3}a$ = 0.246 nm \cite{Nguyen2020}, $E_1^2/B^2 \approx 0.09 $, and  $E_1^2/[2B^2(1+\beta_v^2)] \approx 0.013 \ll 1 $ we can neglected the $E_1$ term and get
\begin{eqnarray}
k_B\Theta_{BG,1} &\approx& \sqrt[3]{\frac{\pi}{6} \frac{1 + \beta_v^2}{1 + \beta_v^5}} \hbar v_{LA} k_F \\
&=& \sqrt[3]{\frac{\left(k_B\Theta_{LA,1}\right)^3 + \beta_v^5\left(k_B\Theta_{TA,1}\right)^3}{1 + \beta_v^5}}.
\end{eqnarray}

\newpage

\newpage
\section{Determinining Bloch-Gr\"{u}neisen temperatures by finding the maximum of $\partial \rho(\mu,T)/\partial T$}

The scattering rate for $\epsilon_{k} = \mu$ as a function of $\mu$ and $T$ can be well described by an analytic expression based on SI approximation\cite{Nguyen2020}
\begin{eqnarray}
\frac{1}{\tau(\mu,T)} &=& 12\Upsilon(\mu) \frac{(v_F k_B T)^4}{\mu^4} \left\{ \frac{1}{v_{LA}^5}
\left[ \frac{2E_1^2}{1+c_1 \alpha_{LA}^3} +\frac {B^2} {1+ c_0\alpha_{LA}^3} \right] + \frac{1}{v_{TA}^5}\frac {B^2} {1+ c_0\alpha_{TA}^3} \right\},\label{tauFT}
\end{eqnarray}
where $\Upsilon(\mu) = \mu^2/4\pi\rho_m\hbar^3v_F^3$, $c_0 = 16.5$, $c_1 = 65.7$, and $\alpha_a = T/\Theta_F^a$ with $k_B\Theta_F^a = 2\hbar v_a k_F$ ($a = LA,\,TA$).  The electrical resistivity in graphene can be approximately written as
\bea
\rho_{\mu}(\mu,T) &=& \frac{1}{\sigma_{\mu}(\mu,T)}= \frac{\pi\hbar^2}{e^2 |\mu|} \frac{1}{\tau(\mu,T)} \nn \\
&=& A Y^4 \left\{ \frac{2E_1^2}{b_d+c_d Y^3} +\sum_{a=LA,TA} \frac {B^2} {d_a + c_a Y^3}  \right\},
\eea
where $Y=k_BT$, $A = \frac{24v_F}{e^2\hbar\rho_m}$, $b_d=8|\mu|^3v_{LA}^5$, $c_d=c_1 v_F^3v_{LA}^2$, $d_a=2|\mu|^3v_{a}^5$, and $c_a=c_0 v_F^3v_{a}^2$.

Taking the second derivative of $\rho_{\mu}(\mu,T)$ with respect to $T$, we get
\be
\frac{\partial^2\rho_{\mu}(T)}{\partial Y^2} = 6 A Y^2 \left\{ \frac{2E_1^2 b_d(2b_d - c_d Y^3)}{(b_d+c_d Y^3)^3}
 +\sum_{a=LA,TA} \frac {B^2d_a(2d_a - c_a Y^3)} {(d_a+ c_a Y^3)^3}\right \},
\ee

Thus, the BG temperature  is a solution of the equation $d^2\rho_{\mu}(\mu,T)/dY^2 = 0$, which we call $\Theta_{BG,2}$.  By omitting the screened deformation potential $E_1$  (which contributes to only about 1\%)  and setting $X \equiv Y^3$, we are left with a solvable quartic equation with respect to $X$,
\begin{eqnarray}
&& \sum_{a=LA,TA} \frac {B^2b_a(2d_a - c_a Y^3)} {(d_a+ c_a Y^3)^3}=0 \nn \\
&\Leftrightarrow& aX^4 + bX^3 + cX^2 + dX + e = 0,
\end{eqnarray}
where
\begin{eqnarray}
a &=&  - (\beta_v + 1)c_{TA}^4 , \\
b &=&  (2\beta_v^4 - 3\beta_v^3 - 3\beta_v + 2)d_{TA}c_{TA}^3, \\
c &=& (- 3\beta_v^6 + 6\beta_v^4 + 6\beta_v^3 - 3\beta_v)d_{TA}^2c_{TA}^2, \\
d &=& (- \beta_v^9 + 6\beta_v^6 + 6\beta_v^4 - \beta_v)d_{TA}^3c_{TA}, \\
e &=& (2\beta_v^9 + 2\beta_v^4)d_{TA}^4.
\end{eqnarray}

\begin{figure}[h]
\includegraphics[width=0.6\linewidth]{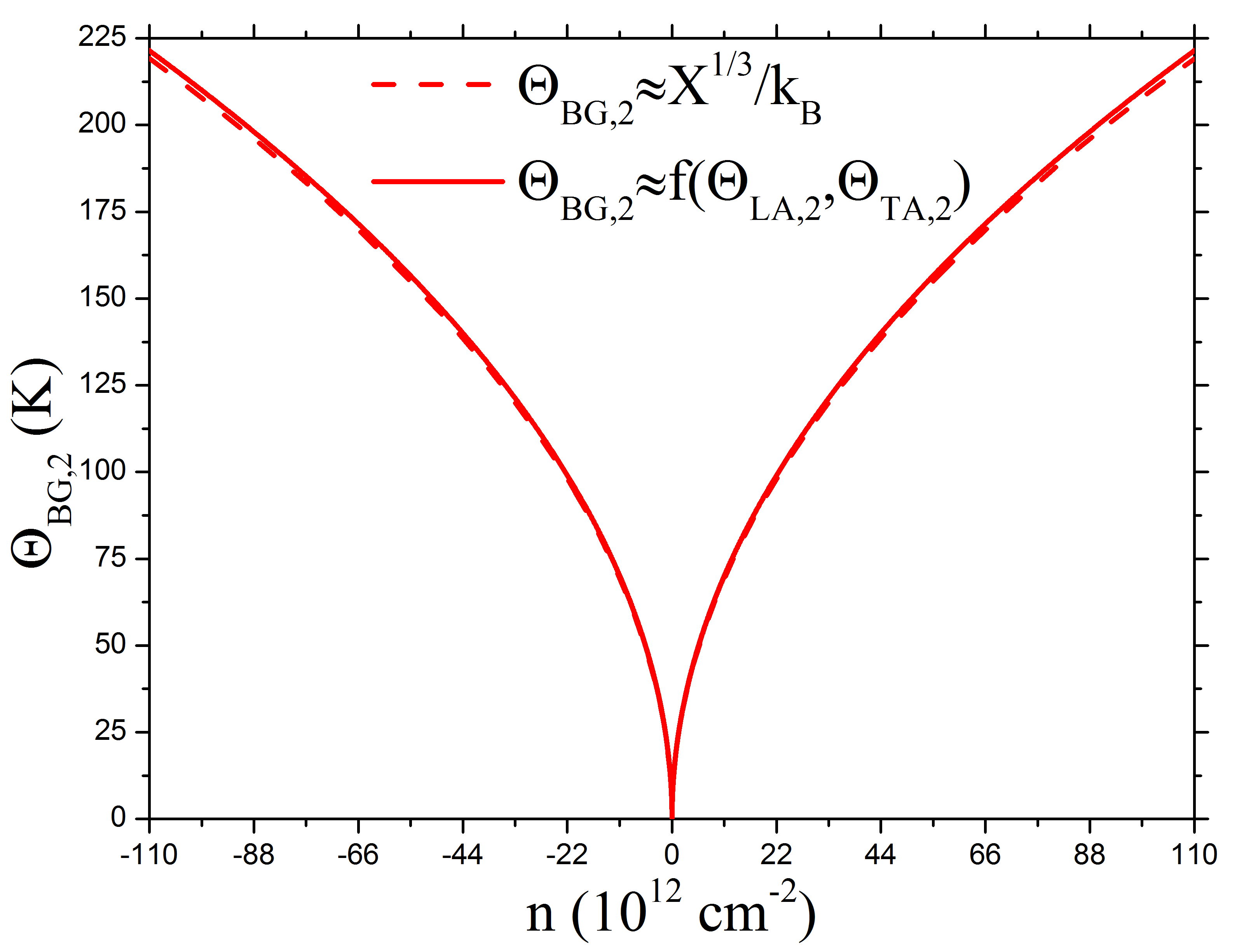}
\caption{\label{sm1} Doping-dependent Bloch-Gr\"{u}neisen temperatures  determined from $d^2\rho_{\mu}(\mu,T)/dT^2 = 0$. Dashed: exact solution given by Eq.~(\ref{exact}). Solid: approximated solution given by   Eq.~(\ref{sim}) with  $f(\Theta_{LA,2},\Theta_{TA,2})=\sqrt[3]{\frac{ \Theta_{LA,2}^3 + \beta_v^2  \Theta_{TA,2}^3}{1 + \beta_v^2}}$. }
\end{figure}

Eliminate the  complex (unphysical), negative, and zero roots, we obtain a single positive  physical solution,
\be
X = \frac{s + \sqrt{s^2 - 4(z_R + t)}}{2} - \frac{{b}}{4{a}} , \label{exact}
\ee
where
\begin{eqnarray}
s &=& \sqrt{2z_R - p}, \\
t &=& - \sqrt{z_R^2 - r}, \\
z_R &=& \sqrt[3]{- \frac{Q}{2} + \sqrt{\frac{Q^2}{4} + \frac{P^3}{27}}} + \sqrt[3]{- \frac{Q}{2} - \sqrt{\frac{Q^2}{4} + \frac{P^3}{27}}} + \frac{p}{6}, \\
P &=& - r - \frac{p^2}{12}, \\
Q &=& - \frac{p^3}{108} + \frac{pr}{3} - \frac{q^2}{8}, \\
p &=& \frac{8ac - 3b^2}{8a^2}, \\
q &=& \frac{b^3 - 4abc + 8a^2d}{8a^3}, \\
r &=& \frac{16ab^2c - 64a^2bd + 256a^3e - 3b^4}{256a^4}.
\end{eqnarray}

Finally, we have
\be
\Theta_{BG,2} \approx \frac{\sqrt[3]{X}}{k_B} \approx \sqrt[3]{\frac{ \Theta_{LA,2}^3 + \beta_v^2  \Theta_{TA,2}^3}{1 + \beta_v^2}}, \label{sim}
\ee
where
\be
 k_B \Theta_{a,2} = \sqrt[3]{\frac{2d_a}{c_a}} = \sqrt[3]{\frac{16}{c_0}} \hbar v_{a} k_F  \approx \hbar v_{a}k_F. \label{TLA2}
\ee
The way we obtain the approximated expression in Eq.~(\ref{sim}) is based on the following argument. In the limit $\beta_v=1$, we should get $\sqrt[3]{X}= k_B \Theta_{a,2}$. Thus, $X$ must take the form $\sqrt[3]{X} = k_B\sqrt[3]{\frac{ \Theta_{LA,2}^3 + \alpha \Theta_{TA,2}^3}{1 + \alpha}}$. We also noticed that $\Theta_{BG,2}\approx 1.35 \Theta_{BG,1}$. Thus, $X$ should be closer to the HT limit than LT limit, and
in the HT  limit we have $\alpha \rightarrow  \beta_v^2$ as implied by Eq.~(\ref{tauHT1}).  In Fig.~ S1, we compare the exact solution given by Eq.~(\ref{exact}) and the approximated solution by   Eq.~(\ref{sim}). It is found that the difference between the two is negligible.

%\bibliography{nvk1ref}% Produces the bibliography via BibTeX.